\documentclass[conference]{IEEEtran}
\IEEEoverridecommandlockouts
\usepackage{balance}
\usepackage{cite}
\usepackage{amsmath,amssymb,amsfonts}
\usepackage{algorithmic}
\usepackage{graphicx}
\usepackage{textcomp}
\usepackage{xcolor}
\def\BibTeX{{\rm B\kern-.05em{\sc i\kern-.025em b}\kern-.08em
    T\kern-.1667em\lower.7ex\hbox{E}\kern-.125emX}}
\begin{document}

\title{Rethinking UX for Sustainable Science Gateways: Orientations from Practice\\
}

\author{\IEEEauthorblockN{Paul C. Parsons}
\IEEEauthorblockA{\textit{School of Applied and Creative Computing} \\
\textit{Purdue University}\\
West Lafayette, USA \\
parsonsp@purdue.edu}
}

\maketitle

\begin{abstract}
As science gateways mature, sustainability has become a central concern for funders, developers, and institutions. Although user experience (UX) is increasingly acknowledged as vital, it is often approached narrowly—limited to interface usability or deferred until late in development. This paper argues that UX should be understood not as a discrete feature or evaluation stage but as a design-oriented perspective for reasoning about sustainability. Drawing on principles from user-centered design and systems thinking, this view recognizes that infrastructure, staffing, community engagement, and development timelines all shape how gateways are experienced and maintained over time. Based on an interview study and consulting experience with more than 65 gateway projects, the paper identifies three recurring orientations toward UX—ad hoc, project-based, and strategic—that characterize how teams engage with users and integrate design thinking into their workflows. These orientations are not a maturity model but a reflective lens for understanding how UX is positioned within gateway practice. Reframing UX as a structural dimension of sustainability highlights its role in building adaptable, community-aligned, and enduring scientific infrastructure.
\end{abstract}


\section{Introduction}

Sustainability has become a defining concern for the science gateways community, as developers, researchers, and funders grapple with how to support platforms beyond the scope of initial grants \cite{gesing_science_2019}. Gateway sustainability is typically framed in terms of funding models, software maintenance, community uptake, and institutional buy-in \cite{gesing_science_2019, gesing_vision_2021}. Yet one crucial dimension remains comparatively under-theorized and under-supported: user experience (UX). UX is often discussed as a property of a system—something a platform “has” or “lacks”—but this view narrows attention to surface interactions and obscures how experience is produced through deeper design, organizational, and infrastructural choices. Here, I frame UX not as a static feature or evaluation outcome, but as a design-oriented perspective for understanding how people and systems co-evolve within the gateway’s broader sociotechnical environment.

While often conflated with usability or interface design, UX in science gateways encompasses a wider set of considerations: how users access, navigate, trust, and integrate platforms into their scientific workflows, and how design decisions, infrastructure, and policy shape those encounters over time. Choices about architecture, support models, onboarding, and responsiveness to user needs all influence a gateway’s capacity to remain viable and adaptable. For science gateways, sustainability also depends on demonstrating impact to funders and building enduring user communities—criteria that align with value-creation expectations in infrastructure grant programs \cite{lawrence_roadmaps_2012}.

In this paper, I argue that UX should not be viewed as a peripheral or cosmetic concern but as a structural element of sustainability planning. I use the term expansively: not as a measure of interface quality, but as a bridge between user-centered design practice and systems-level reasoning. UX, in this sense, is both an orientation for shaping what systems become and a lens for examining how organizational processes influence their long-term viability. Drawing on findings from an interview study and consulting experience with more than 65 gateway teams, I describe how UX emerges not only through design activity, but through timelines, technical constraints, staffing patterns, and institutional decision-making. I identify three recurring orientations toward UX that reflect different ways of engaging with users and embedding design reasoning into gateway development. Rather than a maturity ladder, these orientations provide an interpretive lens and a practical vocabulary for connecting UX effort to sustainable infrastructure.

\section{Rethinking UX: From Interface to Infrastructure}

In science gateways, user experience (UX) is often equated with interface quality or usability testing. While these activities are valuable, such a narrow focus obscures the deeper ways in which users interact with—and depend on—gateway systems. Research in Human–Computer Interaction (HCI) has long emphasized that experience is shaped not only by what appears on a screen but by the surrounding sociotechnical environment: infrastructure, policies, support models, and workflows that together determine how a platform is lived with over time.

Viewed through this broader lens, UX is not merely about visual appeal or ease of use. It concerns how a system fits within scientific practice, how it responds to evolving needs, and how it adapts to change. Star and Ruhleder’s classic account of infrastructure as something that becomes visible only when it misaligns with local practice \cite{star_steps_1994} illustrates this point well. For gateway users, misalignment appears in onboarding bottlenecks, opaque authentication, or absent support—issues that are not interface flaws but systemic frictions that accumulate and shape the overall experience.

Jackson’s notion of repair reinforces this view: the sustainability of technological systems depends less on their initial design than on their capacity to accommodate breakdowns, adaptation, and ongoing maintenance \cite{jackson_rethinking_2014}. Similarly, Kuutti and Bannon urge scholars to understand systems through the lens of practice—what people actually do with them, not simply how they were designed to function \cite{kuutti_turn_2014}. Together, these perspectives locate UX not in isolated interactions but within the institutional and infrastructural conditions that make interaction possible in the first place.

This reframing invites a shift from thinking about UX as a discrete feature to be optimized to seeing it as a perspective for reasoning about infrastructure. It asks gateway teams not only “Is this usable?” but also “How is this system supported, adapted, and sustained?” Such a stance also redefines what we mean by “users.” Conventional approaches cast users as passive recipients whose experience is determined by design. In practice, users actively interpret, adapt, and repurpose technologies to meet their goals \cite{mccarthy_technology_2004}. Infrastructure-focused research likewise shows that systems are experienced through breakdowns, improvisations, and workarounds \cite{star_steps_1994, jackson_rethinking_2014, koopman_work-arounds_2003}. In science gateways, this often takes the form of creative adaptations—using features in unintended ways or developing scripts and workflows to fill gaps. Recognizing this adaptive capacity underscores the need for a UX perspective that attends not only to usability and interface design but to how systems are inhabited and maintained in real-world contexts.

To treat UX as a systems-level concern is therefore to move beyond a purely evaluative mindset. UX should function not only as a diagnostic tool for identifying issues but as a generative mode of inquiry—a way of imagining, exploring, and shaping more sustainable futures for scientific infrastructure.

\section{From Evaluation to Imagination: A Design-Oriented View of UX}

In many gateway projects, UX is approached primarily as a form of evaluation. Usability testing or heuristic reviews are often conducted late in the development process, once core infrastructure and features have already been implemented. While such evaluation can be valuable, it is inherently backward-looking—it asks whether something that has already been built works well for users.

A design-oriented approach offers a different mindset. Rather than treating UX as a final-stage check, it positions UX as a starting point for exploring what a gateway could become. This orientation emphasizes generativity over diagnosis \cite{hutchison_user_2006}. It encourages teams to engage in activities like brainstorming, sketching, and prototyping—not to refine what already exists, but to imagine alternatives and surface latent needs \cite{Fallman2003}. These practices are not merely about aesthetics or functionality; they are tools for inquiry. They help teams surface assumptions, reframe problems, and open up new directions for what the system might support.

In this way, a design mindset supports thinking in terms of opportunity, not just usability. It invites teams to ask, “What if?” and “What else?” before committing to a technical path. Even lightweight practices—such as mapping user flows on a whiteboard or sketching divergent concepts—can clarify priorities and better align design decisions with the experiences of users.

Adopting this orientation does not mean abandoning evaluation. Rather, it reframes UX as both a generative and reflective practice. Design methods complement evaluative ones by helping teams shape the problem space before narrowing in on solutions. This shift supports more sustainable outcomes by fostering gateways that are not only usable, but adaptable, contextually grounded, and responsive to evolving needs.

\section{Grounded Reflections on UX Practice}

The reflections offered in this paper are grounded in two sources of insight: a prior interview study with science gateway teams \cite{parsons_what_2024} and experiences from consulting with over 65 gateway projects. The interview study examined how teams engage with UX, including the challenges they face and the strategies they use to integrate user input into design and decision-making. Our research team conducted semi-structured interviews with gateway developers, principal investigators, and team members, and used thematic analysis to identify cross-cutting patterns of practice. These findings are reinforced by consulting work conducted through SGCI and SGX3, which involved short-term engagements with gateway teams at various stages of development and maturity. Earlier work drawing on this consulting activity \cite{parsons_common_2020} identified recurring interface-level usability issues—such as weak visual hierarchy and navigation cues—that hinder user adoption. The present paper extends that line of inquiry from usability problems to the organizational orientations that give rise to them, offering a systems-level perspective on how UX is framed within gateway practice.

Across both sources, I found that UX was frequently treated as a late-stage concern—addressed only after core functionality had been implemented. Even when teams valued UX, they often lacked the capacity, vocabulary, or organizational support to fully integrate it into their processes. As one participant explained, ``That's where some of our usability problems came from—we had to do it in the time we had, not necessarily how we would've liked to. It was driven by other priorities.''

In many cases, constraints on UX were embedded in infrastructural and institutional decisions. For example, reliance on legacy systems or rigid funding structures limited the extent to which teams could make design changes, even when usability issues were well known. One team described inheriting an authentication system that users found confusing but that could not be modified without extensive reengineering. Another participant noted that project timelines driven by grant deadlines left little room for iteration: ``We knew some of the workflows weren’t ideal, but we didn’t have time to fix them.''

UX was also shaped by the presence—or absence—of support infrastructure and community engagement practices. Some teams had no dedicated channels for gathering user feedback. Others described internal uncertainty about who was responsible for implementation, leading to what one participant called a ``handoff gap'' between usability recommendations and actual development. As another participant reflected, ``The usability group [our consulting team] brought us together and got us on the same page about what we were even trying to do. It helped clarify our own internal confusion before we could even think about the users.''

These examples suggest that UX challenges in science gateways are not simply design oversights—they are symptoms of deeper structural dynamics, shaped by how projects are funded, staffed, and organized. UX, in this light, is a form of infrastructural labor that contributes directly to a gateway’s sustainability, adoption, and impact. Rather than viewing UX as a peripheral task or aesthetic concern, I argue for treating it as integral to the broader sociotechnical system in which gateways are developed and used.

\section{Patterns of UX Engagement in Gateway Projects}

The prior section highlighted how UX in science gateways is often constrained by timelines, infrastructure, staffing, and institutional structure. While these constraints vary, they produce recognizable patterns in how gateway teams approach UX. Based on interview data and consulting experience, I have identified three recurring orientations that reflect different ways of thinking about UX in practice. These orientations are not formal maturity levels, nor do they represent a developmental sequence. Instead, they offer a vocabulary for characterizing how UX tends to be positioned within gateway projects—from reactive, ad hoc efforts to more strategic and embedded approaches. By articulating these orientations, I aim to support reflection among teams and help make visible the often implicit assumptions that shape how UX is understood and practiced.

These orientations are not rigid categories or evaluative scores, but interpretive tools. Gateway teams may exhibit a mix of these patterns, often shaped by local constraints and histories. The strategic orientation, in particular, should be understood as a direction of travel rather than a current standard—one that reflects what is possible when UX is integrated as both a practice and a mindset.

\subsection{Ad Hoc Orientation}

In this orientation, attention to UX tends to be informal, episodic, and reactive. Gateway teams may acknowledge usability issues or barriers to adoption, but these are typically addressed through quick fixes or workarounds—often in response to user complaints or issues recognized internally. There are few, if any, dedicated resources for UX work, and no structured processes for gathering feedback, exploring alternatives, or evaluating impact. Design decisions are driven primarily by technical feasibility or internal assumptions rather than by systematic engagement with users or UX methods. In many cases, UX is treated as a concern to be addressed after core infrastructure is already in place—if time allows. While this approach may lead to functional systems, it limits the team’s ability to anticipate problems, engage users meaningfully, or identify new opportunities for growth and sustainability.

\subsection{Project-Based Orientation}

Teams in this orientation recognize the value of UX and incorporate it more explicitly into their work, often within the scope of specific grants or time-bounded development cycles. They may conduct usability evaluations, run surveys, or engage users through workshops and feedback sessions. These efforts can meaningfully improve interfaces and clarify user needs. However, they are often scheduled late in the development process and framed primarily as evaluation tasks—focused on identifying issues in a system that is already largely defined. Generative design practices, such as early-stage sketching, prototyping, or collaborative problem framing, are rare or inconsistently applied. As a result, UX may inform refinements but seldom shapes the overall direction of the project. Without institutional support or continuity between project phases, UX efforts may stall, and valuable insights can be lost over time.

\subsection{Strategic Orientation}

In this orientation, UX is treated as an ongoing, cross-cutting concern that informs both design decisions and long-term sustainability planning. Teams proactively incorporate user research, design exploration, and iterative feedback into their development processes. UX is not confined to interface tweaks or late-stage evaluations—it is used to ask generative questions, explore possibilities, and shape infrastructure in ways that align with evolving user needs. Even lightweight practices such as sketching workflows, mapping user journeys, or co-designing prototypes are used strategically to frame and reframe the problem space.

While elements of this orientation are visible in some projects—particularly those with strong UX champions or previous consulting engagements with our teams—it remains somewhat rare in practice. Few gateway teams fully realize this orientation, often due to limitations in staffing, funding, or organizational culture. Nonetheless, it serves as an aspirational view that can help guide more intentional integration of UX thinking over time. It reflects a mature understanding of UX as both a design practice and a strategic asset—one that supports more sustainable, adaptable, and community-aligned infrastructure.

\subsection{Synthesizing the Orientations}

Taken together, these orientations illuminate not simply different levels of UX investment but distinct logics of sustainability. Ad hoc efforts often trade flexibility for increasing ``UX debt,'' as short-term fixes create inconsistencies that are costly to maintain. Project-based approaches introduce structure and yield visible gains, yet their episodic nature limits organizational learning—design knowledge is easily lost when funding cycles or staff change. Strategic approaches, by contrast, distribute UX responsibility across governance and development, embedding feedback, design systems, and continuity mechanisms that amortize UX debt over time. Viewing the orientations comparatively thus reveals how UX practices operate as sustainability strategies: each reflects a different coupling between design, infrastructure, and institutional support.

\section{Looking Ahead}

This paper has argued for a broader conception of UX in science gateways—one that moves beyond interface-level concerns and evaluation-driven mindsets, and toward a design-oriented view that treats UX as integral to sustainability. By reframing UX as a lens for reasoning about infrastructure, workflows, and engagement, we can better recognize how gateways are lived with, adapted, and sustained over time.

For mature gateways constrained by legacy systems, UX improvements can also proceed incrementally. Establishing a lightweight ``UX debt'' list, mapping user journeys against backstage workflows, or consolidating recurring interface elements into simple design patterns can yield tangible gains without large refactoring. Such retrofits reframe UX work as infrastructural maintenance—small, continuous adjustments that sustain usability over time.

The three orientations described here offer one way to surface common patterns in UX practice. While many teams operate under constraints that make sustained UX engagement difficult, there are promising signs that a more strategic approach is possible. For example, in one engagement, a gateway team worked with SGCI consultants to conduct early-stage user research, prototype alternative workflows, and revise their onboarding process before development began. These generative activities were modest in scope but led to clearer internal alignment and more confident design decisions. While such examples remain somewhat rare, they suggest what is possible when UX is treated as a shared responsibility and a strategic asset.

Looking ahead, I hope this work encourages teams to reflect not only on their UX practices, but on how those practices relate to broader structures of funding, team organization, and long-term planning. Treating UX as infrastructure—as something that supports, enables, and evolves with the system—can help ensure that science gateways remain usable, adaptable, and valuable over time.

% \section{Conclusion and Call to Action}

% As the science gateways community continues to invest in long-term sustainability, there is growing recognition that technical robustness, funding stability, and community engagement must go hand in hand. In this paper, we have argued that user experience (UX) should be considered a foundational part of this equation—not only in terms of interface design, but as a reflection of how infrastructure decisions, project planning, and support strategies shape the user’s relationship to the gateway.

% Drawing from interviews and consulting experiences across more than 65 gateway projects, we observed that while many teams care deeply about serving their users, they often face structural and conceptual barriers to embedding UX more fully into their work. The UX maturity model we propose offers a flexible scaffold for reflecting on current practices and planning for more strategic integration over time.

% As funding agencies and institutions look for models of sustainable scientific infrastructure, we see an opportunity to broaden how UX is understood and supported. Rather than treating UX as a feature to be evaluated or retrofitted, I argue for treating UX as a way of reasoning about design, maintenance, and long-term use in complex systems. This perspective links user experience to sustainability—not only through better interfaces, but through better systems thinking.

\balance
\bibliographystyle{IEEEtran}
\bibliography{refs}
\end{document}